\documentclass[12pt,titlepage]{article}
\usepackage[utf8]{inputenc}
\usepackage[left=1in,top=1in,right=1in,bottom=1in]{geometry}
\usepackage{graphicx}
\usepackage{amsmath}
\usepackage{siunitx}
\usepackage[utf8]{inputenc} 
\usepackage{amsfonts}       
\usepackage{nicefrac}       
\usepackage{microtype}      
\usepackage{amsmath, amsfonts,bm,amssymb,indentfirst,mathtools,bbm}
\RequirePackage[colorlinks,citecolor=blue,urlcolor=blue]{hyperref}
\usepackage[noend]{algpseudocode}
\usepackage{booktabs}
\usepackage{subcaption}
\usepackage{algorithmicx,algorithm,enumerate,accents}
\usepackage{color,xcolor}
\usepackage{microtype}
\usepackage{threeparttable}
\usepackage{natbib}
\usepackage{authblk}
\usepackage{setspace}
\usepackage{blkarray}

\allowdisplaybreaks

\newcommand\pkg[1]{\texttt{#1}}
\let\proglang=\textsf

\title{Robust Community Detection for Noisy Networks with Covariates:
Application to Functional Brain Networks}

\author[1]{Zeyu Hu}
\author[2]{Frederick H.\ Xu}
\author[3]{Suprateek Kundu}
\author[4]{Yize Zhao}
\author[5]{Li Shen}
\author[1]{Jun Yan}
\author[1,*]{Wenrui Li}
\affil[1]{Department of Statistics, University of Connecticut,
	Storrs, CT 06269}
\affil[2]{Department of Bioengineering, University of Pennsylvania,
	Philadelphia, PA 19104}
\affil[3]{Department of Biostatistics, University of Texas MD Anderson
	Cancer Center, Houston, TX 77030}
\affil[4]{Department of Biostatistics, Yale University,
	New Haven, CT 06520}
\affil[5]{Department of Biostatistics, Epidemiology and Informatics,
	University of Pennsylvania, Philadelphia, PA 19104}
\affil[*]{Correspondence:
	\href{mailto:wenrui.li@uconn.edu}{Wenrui Li}}

\begin{document}
\maketitle

\begin{abstract}
Community detection is fundamental to understanding the modular 
organization in functional brain networks, yet noise in 
neuroimaging-derived networks and auxiliary node-level covariates pose
critical challenges. Existing methods typically either assume networks are
noise-free or ignore covariate information. We propose a Bayesian framework for
recovering a shared latent community structure from multiple noisy network
realizations and auxiliary covariates. The model combines a
degree-corrected stochastic block model for the latent network, a
block-structured noise model linking noisy observations to latent edges, and
a covariate cluster model for node-level attributes. This specification allows
anatomical or functional attributes of regions of interest to contribute information when
network signals are weak or sparse. We develop an efficient Markov chain
Monte Carlo algorithm for posterior sampling and select the number of
communities using the widely applicable information criterion, avoiding prior
specification of this quantity. Simulation studies demonstrate improved
community recovery relative to existing methods across varying noise levels,
covariate signal strengths, and numbers of noisy networks, with larger gains
when network noise is moderate to high or only a small number of noisy
networks is available. Applications to functional brain networks from the
Alzheimer's Disease Neuroimaging Initiative and the Human Connectome Project
identify biologically interpretable structures and capture disease-related
reorganization and individual-level variation.

\bigskip

 
\end{abstract}

\section{Introduction}
\label{sec:intro}

The human brain is a highly complex system with organized
structural and functional architectures~\citep{bullmore2009complex, 
rubinov2010complex, sporns2013structure, sporns2016networks}. Recent 
advances in magnetic resonance imaging (MRI), particularly functional 
MRI (fMRI), have enabled in vivo characterization of large-scale brain 
organization and connectivity~\citep{biswal1995functional, 
hagmann2008mapping, smith2013resting, sporns2005human}. A fundamental 
property of brain organization is its modular structure, where
groups of regions of interest (ROIs) form communities supporting
specialized functions and interactions~\citep{meunier2009hierarchical,
meunier2010modular}. Functional brain network analysis has emerged 
as a central paradigm for quantifying and interpreting this modular 
organization~\citep{power2011functional, yeo2011organization}. This
approach has been widely used to characterize individual variation in
brain organization~\citep{gordon2017precision, mueller2013individual} 
and changes in functional network organization associated with
neurodegenerative  diseases such as Alzheimer's disease
(AD)~\citep{xu2024topology, seeley2009neurodegenerative,
supekar2008network}.

To characterize modular organization, community detection plays a 
central role in functional brain network analysis. In this framework, 
the brain is represented as a network, where nodes correspond to ROIs 
and edges encode the statistical dependencies between regional 
signals~\citep{logothetis2002neural, ogawa1990brain}. Identifying 
communities provides a data-driven characterization of large-scale 
brain systems, capturing functionally specialized yet 
interacting subnetworks. A wide range of community detection methods have
been developed, including modularity-based methods~\citep{blondel2008fast,
newman2006modularity}, spectral clustering~\citep{ng2002spectral,
vonluxburg2007tutorial}, and stochastic block models and their
extensions~\citep{airoldi2008mixed, holland1983stochastic}.

Despite the widespread use of these community detection methods, 
several challenges remain. At the individual level, functional 
connectivity is not directly observed but estimated from neuroimaging 
data, introducing both measurement and estimation 
uncertainty~\citep{murphy2013resting, power2012spurious, 
smith2011network}. Consequently, repeated measurements from the same 
individual can yield noisy network realizations, and failure to 
account for such noise may reduce the accuracy of community detection.
At the population level, a central goal is to 
recover a shared organizational structure from a collection of 
individual networks, which can be viewed as noisy realizations of an 
underlying population level template. This problem is particularly 
relevant in AD studies, where comparing community structure across 
disease stages provides insights for characterizing progressive 
network disruption and identifying deviations from normative 
functional organization~\citep{xu2024caudal, sheng2024novel}.

In addition, neuroimaging studies often collect auxiliary ROI-level 
covariates, including spatial, anatomical, and functional attributes 
of brain regions~\citep{binkiewicz2017covariate, glasser2016multi, 
schaefer2018local}. Such information complements network topology and
can improve community recovery, particularly when 
network signals are sparse or weak~\citep{deshpande2018contextual, 
yan2021covariate}. Integrating these covariates with network data, 
while simultaneously accounting for network noise at both the 
individual and population levels, remains a critical challenge.

Existing approaches address these challenges only partially. Classical 
community detection methods rely solely on network topology and assume the
observed network is noise free. Covariate-assisted approaches
\cite[e.g.,][]{binkiewicz2017covariate,
hu2024network, louit2025calfsbm, shen2025bayesian} use node-level 
covariates but do not account for network noise. Recent methods that account
for network noise \cite[e.g.,][]{hu2024joint} do not incorporate node
covariate information. To the best of our knowledge, no existing community
detection framework simultaneously accounts for network noise and leverages
node-level covariate information.

Our contribution in this paper is to provide a Bayesian community detection
framework that jointly accounts for network noise and leverages nodal
covariate information, without requiring prior knowledge of the number of
communities. The method treats the true network as latent and has three key
components. We employ a
degree-corrected stochastic block model~\citep[DC-SBM,][]{peng2016bayesian}
to characterize community structure in the underlying network. The DC-SBM 
is integrated with a block-structured network noise model
\citep{le2018estimating} that relies on noisy networks to inform latent
communities. These components are combined with a covariate cluster model,
adapted from existing approaches
\citep{jin2016influential,jin2017phase,shahbaba2009nonlinear},
to incorporate node-level covariate information for community detection.
This joint specification uses the complementary information in noisy networks
and node covariates to improve estimation of latent community structure. For
brevity, we refer to the proposed approach as RCDC, Robust Community Detection
for noisy networks with Covariates. We evaluate RCDC through simulation
studies with varying network noise levels and numbers of noisy networks, and
through applications to functional brain networks from Alzheimer's Disease
Neuroimaging Initiative (ADNI) study~\citep{mueller2005alzheimer} and Human
Connectome Project~\citep[HCP,][]{van2013wu}.

The organization of this paper is as follows. In 
Section~\ref{sec:method} we present the proposed methodology 
and the Markov chain Monte Carlo (MCMC) algorithm. Section~\ref{sec:sim} 
reports the practical performance of our method through simulation studies.
In Section~\ref{sec:real_data} we apply our method to functional brain
networks from the ADNI and HCP studies. Finally, we conclude in 
Section~\ref{sec:disc} with a discussion of future directions for this 
work. The details of the computation framework are relegated to Web 
Appendix A.

\section{Methodology}
\label{sec:method}

We present our proposed model first and then the associated
computation framework.

\subsection{Model Specification}
\label{sec:spec}

Let $G=(V,E)$ denote the true underlying undirected graph on $n$ nodes,
with vertices $V=\{1,\dots,n\}$ and a set $E$ of edges, where
elements of $E$ are unordered pairs $\{i,j\}$ of distinct vertices 
$i,\ j\in V$. Suppose we have data of $M$ noisy graphs, denoted by 
$G^{(m)}=(V,E^{(m)})$ for $m=1,\dots,M$, where we implicitly assume
that the vertex set $V$ is known. Denote the adjacency matrix of $G$ 
by $\bm A=(A_{ij})_{n\times n}$ and that of $G^{(m)}$ by 
$\tilde {\bm A}^{(m)}=(\tilde A_{ij}^{(m)})_{n\times n}$.
Hence $A_{ij} = 1$ if there is a true edge between the $i$-th 
vertex and the $j$-th vertex, and 0 otherwise, while 
$\tilde A_{ij}^{(m)} = 1$ if there is an edge in the noisy graph
$G^{(m)}$ between the $i$-th vertex and the $j$-th vertex, and 0
otherwise. We assume throughout that $G$ and $G^{(m)}$, $m=1,\dots,M$,
are simple, i.e., that they possess neither multi-edges nor 
self-loops. In addition, assume that we have covariate information 
$\bm x_i \in \mathbb R^p$ for each node $i=1,\dots,n$. Our goal is 
to detect communities by leveraging this nodal covariate information 
while accounting for noise in $G^{(m)}$, $m=1,\dots,M$.

We model the structure in $\bm A$ through one of the most commonly
used network community models, DC-SBM. 
Let $K$ be the number of communities. Note that our method does not
require prior knowledge of $K$, and the criterion for selecting
$K$ is described in Section~\ref{sec:mcmc}. Define a symmetric matrix 
$\bm B=(B_{kl})_{K \times K}$, a node membership vector $\bm{z} = 
(z_1,\dots,z_n)\in [K]^n$ where $[K]=\{1,\dots,K\}$ for any $K\in 
\mathbb N$, and a vector of node degree-correction parameters 
$\bm\theta=(\theta_1, \dots,
\theta_n)^{\top}$. For all $1\le i<j\le n$, we assume
\begin{align}
	A_{ij}\mid \bm z,\bm B,\bm\theta \sim \mathrm{Bern}\left(
    \mathrm{logit}^{-1}\left( B_{z_i z_j} + \theta_i+\theta_j \right)
    \right),
\end{align}
where Bern$(\cdot)$ denotes the Bernoulli distribution. On the logit 
scale, $\bm B$ captures within and between community connectivity. 
Following~\citet{peng2016bayesian}, we impose an identifiability 
constraint by setting $B_{11}=\cdots=B_{KK}=0$, yielding
\begin{align}
	A_{ij} \mid \bm z,\bm B,\bm\theta \sim \mathrm{Bern}\biggl(
	\mathrm{logit}^{-1}\Bigl\{ B_{z_i z_j} \mathbb{I}
	\left( z_i\neq z_j \right) + \theta_i+\theta_j \Bigr\}\biggr),
    \label{DCSBM}
\end{align}
where $\mathbb{I}(\cdot)$ denotes the indicator function. We assign the 
following prior to $\bm \theta$ and $\{B_{kl}: 1\leq k< l \leq K\}$:
\begin{align}
	\left( \{B_{kl}: 1\leq k< l \leq K\} , \bm \theta^\top\right)^\top & \sim 
	\mathbb{I}\left(\{B_{kl}: 1\leq k< l \leq K\} \leq 0\right) \cdot 
    \mathcal{N} (0,\sigma^2 \bm I_{n+K(K-1)/2}),
\end{align}
where $\mathcal N(\cdot)$ denotes the Gaussian distribution and 
$\bm I$ denotes the identity matrix. The constraint $B_{kl}\leq 0$ 
encourages edge densities between communities to be no greater than 
those within communities, which is essential for community detection.

Adapting the block-structured network noise assumption of
\citet{le2018estimating}, we specify the
model for $\tilde {\bm A}^{(m)}$, $m=1,\dots,M$, as follows: for $1
\leq i<j\leq n$,
\begin{align}
\begin{split}
	\Pr \left(\tilde A_{ij}^{(m)}=1 \mid 
	A_{ij}=0,\ z_i=k,\ z_j=l\right) = \alpha_{kl}, \\
	\Pr \left(\tilde A_{ij}^{(m)}=0 \mid 
	A_{ij}=1,\ z_i=k,\ z_j=l\right) = \beta_{kl}.    
    \label{eq_noisy_network}
\end{split}
\end{align}
Here $\alpha_{kl}$ and $\beta_{kl}$ 
represent the false positive and false negative rates, respectively,
for node pairs whose endpoints belong to communities
$k$ and $l$. We assume symmetry, i.e., $\alpha_{kl}=\alpha_{lk}$ and
$\beta_{kl}=\beta_{lk}$, for all $k\neq l, k,l=1,\dots,K$. By 
convention, we assign the beta distributed priors to the error rates:
\begin{align}
	\alpha_{kl} \sim \text{Beta}(a_\alpha,b_\alpha),\ \beta_{kl}  \sim
    \text{Beta}(a_\beta,b_\beta),\ 1\leq k\leq l\leq K, 
\end{align}
where $\text{Beta}(\cdot)$ denotes the Beta distribution.

We model the covariates $\bm x_i$ using a standard cluster 
model~\citep{jin2016influential,jin2017phase}, and adapt the conditional
independence assumption in \citet{hu2024network}. Specifically, the
covariates $\bm x_i$ are independently distributed as
\begin{align}
	\bm x_i \mid \bm z \sim F_{z_i},
\end{align}
where $F_k$ denotes a distribution for community $k$, $k\in [K]$. When $p$
is moderate to large, estimation of the joint distribution is both 
statistically and computationally challenging. We adapt the approach of
\citet{shahbaba2009nonlinear} to address this. Specifically, we assume the 
covariates are independent within components, i.e., $p(\bm 
x_i|z_i)=\prod_{r=1}^p p(x_{ir}|z_i)$, where $x_{ir}$ denotes the $r$th 
covariate of node $i$. We provide details below for normal and categorical
distributions, which are commonly used. The model can also be extended to 
accommodate other distributions. For the normal distribution case, we specify
\begin{align}
\begin{split}
	x^\text{cont}_{i r}\mid z_i=k,\mu_{k r},\tau^2_{k r}
	&\sim \mathcal N(\mu_{k r},\tau^2_{k r}),  \\
    	\mu_{k r}\mid \tau^2_{k r} &\sim
	\mathcal N(\mu_{0 r},\,\iota_{0 r}\tau^2_{k r}),\\
	\tau^2_{k r} &\sim \mathcal{IG}(a_{\tau r},b_{\tau r}),
\end{split}
\end{align} 
where $x^\text{cont}_{i r}$ denote a continuous covariate of node $i$, and 
$\mathcal{IG}(\cdot)$ denotes the inverse 
gamma distribution. Let $x^\text{cat}_{i s}$ be a categorical covariate of node 
$i$ and $x^\text{cat}_{i s}\in\{1, \dots,C_s\}$. Denote the corresponding vector
of category probabilities in community $k$ by 
$\bm\xi_{k s}=(\xi_{k s 1}, \dots, \xi_{k s C_s})$, i.e.,  
\begin{align}
	\Pr \left(x^\text{cat}_{i s}=c\mid z_i=k,\bm\xi_{k s}\right)=\xi_{k s 
    c},\ c=1, \dots,C_s.
	\label{cat_P}    
\end{align}
We assign the following prior to $\bm\xi_{k s}$:
\begin{align}
	\bm\xi_{k s}\sim \text{Dir}(\gamma_{s} \bm 1_{C_s}),
\end{align}
where $\text{Dir} (\cdot)$ denotes the Dirichlet distribution and $\bm 1$ denotes
the all-ones vector.

Following \citet{peng2016bayesian}, we assign a constrained multinomial 
distributed prior to $\bm z$ :
\begin{align}
	p(\bm z \mid \bm \pi )&\propto \prod_{k=1}^K \mathbb{I}(N_k>1) 
	\prod_{i=1}^n \pi_k^{\mathbb{I}(z_i=k)},
\end{align}
where $\bm\pi=(\pi_1, \dots,\pi_K)$ is a vector of prior probabilities over
$K$ labels, and $N_k = \sum_{i=1}^n \mathbb{I}(z_i = k)$ 
denotes the size of community $k$. The constraint $N_k>1$ is imposed to 
ensure model identifiability. In addition, we assign a Dirichlet 
distributed prior to $\bm\pi$:
\begin{align}
	\bm\pi \sim \text{Dir} \left(\frac{d}{K}\bm 1_K\right).
\end{align}

\subsection{Computation}
\label{sec:mcmc}

We develop an efficient MCMC algorithm for sampling the parameters in 
our model. To facilitate the sampling, we augment P$\acute{\text{o}}$lya-Gamma 
latent variables for Bernoulli distributed variables
$A_{ij}$, $1\leq i<j\leq n$.
All parameters are sampled using the Gibbs sampler. We adopt
the label remapping approach in~\citet{peng2016bayesian} to address the 
label identifiability issue, and motivated by~\citet{le2018estimating}, 
propose an initialization strategy to accelerate MCMC convergence. 
Details of the computation framework can be found in the
Web Appendix A.

Following similar works~\citep{merkle2019bayesian,tan2024bcclong,louit2025calfsbm}, we 
choose the number of communities $K$ by the widely applicable information
criterion~\citep[WAIC,][]{watanabe2010asymptotic},
\begin{align}
    \text{WAIC}=&-2\sum_{m=1}^M\sum_{1\leq i<j\leq n} \biggl\{ \log\Bigl( 
    \mathbb{E}_{\mathrm{post}}\bigl[
    p(\tilde A_{ij}^{(m)} \mid \Theta)\bigr]\Bigr) -
    \mathrm{Var}_{\mathrm{post}}\bigl[
    \log p(\tilde A_{ij}^{(m)} \mid \Theta)
    \bigr]\biggr\}
    \nonumber\\
    &-2\sum_{i=1}^{n} \biggl\{\log\Bigl(
    \mathbb{E}_{\mathrm{post}}\bigl[
    p(\bm x_i \mid \Theta)
    \bigr]\Bigr) -
    \mathrm{Var}_{\mathrm{post}}\bigl[
    \log p(\bm x_i \mid \Theta)
    \bigr]
    \biggr\},
\end{align}
where $\Theta$ denotes all model parameters,  
$\mathbb{E}_{\mathrm{post}}$ and $\mathrm{Var}_{\mathrm{post}}
$ denote the expectation and variance under the posterior 
distribution of $\Theta$, respectively. 
We choose $K$ by minimizing the WAIC.

We develop an open-source and user-friendly \proglang{R} package,
\pkg{rcdc}, for implementing the proposed method. The package is publicly 
available on GitHub at
\href{https://github.com/zyhu888/rcdc}{https://github.com/zyhu888/rcdc}.

\section{Simulation Studies} 
\label{sec:sim}

We conducted simulations to evaluate the performance of our
model in comparison with several existing methods under a range of
settings, varying the number of noisy networks and their noise levels, as 
well as the strength of the network structure and the signal in the node
covariates.

\subsection{Data Generation}

We generated the true network from the DC-SBM in (\ref{DCSBM}) with $B_{kl}
= b$ for $k \neq l,\ k,l =1,\dots,K$, and $\theta_i \mid z_i = k \sim 
\mathcal{N}(\mu^{\theta}_k, 0.1)$, $i=1,\dots,n$. We set $n = 200$ nodes 
with $K = 8$ equally sized communities. The parameters 
$(\mu^{\theta}_1,\dots,\mu^{\theta}_8)$ were fixed at $(0.5,  0.3, 
0.1,  -0.1,  -0.3,  -0.5, -0.7,  -0.9)$. The parameter $b$ controls the 
magnitude of disparity between within- and between-community connectivities. 
We considered two levels of the network structure: a strong community structure
with $b = -2$ and a weak community structure with $b = -1.7$.

Noisy networks were generated from model (\ref{eq_noisy_network}), with 
within- and between-community false positive and false negative 
probabilities given by
\begin{align*}
	\alpha_{kl} =
	\begin{cases}
		\rho_{\alpha}, & \text{if } k = l, \\
		\rho_{\alpha} \zeta_{\alpha}, & \text{if } k \neq l,
	\end{cases}\hspace{1cm} 
    \beta_{kl} =
	\begin{cases}
		\rho_{\beta}, & \text{if } k = l, \\
		\rho_{\beta} \zeta_{\beta}, & \text{if } k \neq l.
	\end{cases}
\end{align*}
Here, $\rho_{\alpha}$ and $\rho_{\beta}$ control the overall noise level,
and $\zeta_{\alpha}$ and $\zeta_{\beta}$ capture the relative magnitude of 
within- versus between-community noise. We considered various values of 
these parameters to examine the effect of network noise on the performance 
of our method. In addition, we varied the number of noisy networks to assess
its impact.

We considered a mix of continuous and categorical covariates, with a total
of $p=10$ covariates. The continuous covariates $\bm{x}^\text{cont}_i \in \mathbb
R^8$ were generated from a multivariate normal distribution (MVN): 
$\bm{x}^\text{cont}_i|z_i = k \sim \mathrm{MVN} (\bm{\mu}_k, \tau^2 \bm{I}_8)$,
where $\bm{\mu}_k = (\mu_{k 1}, \dots, \mu_{k 8})$. The cluster centers 
$\bm{\mu}_k$ were positioned at eight vertices of a 7-simplex centered at the
origin with fixed edge length 10. The parameter $\tau$ controls the strength
of the continuous covariate signal. We further considered two categorical
covariates, where each covariate $x^\text{cat}_{is}\in\{1,\dots,C_s\}$ was
generated according to (\ref{cat_P}) with $C_s=8,\ \xi_{k s c} = \xi$ if $c =
k$, and $\xi_{k s c}= (1-\xi)/(C_s - 1)$ for all $c \neq k$. The parameter
$\xi$ controls the strength of the categorical covariate signal. We
considered two levels of covariate signal: a strong case with $\tau = 3$ and
$\xi = 0.7$, and a weak case with $\tau = 3.5$ and $\xi = 0.6$.

Clustering accuracy was evaluated using the Adjusted Rand Index \citep[ARI,]
[]{rand1971objective}, which quantifies the agreement between the estimated
and true community assignments. We set fairly uninformative priors: 
$\sigma=10$ for $(\bm B,\bm\theta)$; $\alpha_{kl}, \beta_{kl} \sim 
\text{Beta}(1,1)$; $\mu_{0 r}=0$ and $\iota_{0 r}=1$ for $\mu_{k r}$; 
$\tau^2_{k r}\sim \mathcal{IG} (1,0.01)$; $\gamma_s=1$ for $\bm\xi_{ks}$;
and $d=1$ for $\bm\pi$. We ran MCMC chains of 1000 samples. 
We discarded the first 500 samples as burn-in and assessed convergence
using trace plots.

\subsection{Methods to Compare}

Our method was compared with several approaches. We included an existing
community detection method for noisy networks, the 
group-based binary mixture (GBM) modeling approach~\citep{hu2024joint}. 
We also considered covariate-assisted community detection methods, namely the 
covariate-assisted spectral clustering (CASC) 
algorithm~\citep{binkiewicz2017covariate}, the covariate-assisted spectral 
clustering on ratios of eigenvectors (CASCORE)~\citep{hu2024network}, the 
Bayesian community detection for networks with covariates 
(BCDC)~\citep{shen2025bayesian}, and the covariate-assisted latent factor 
stochastic block model (CALF-SBM)~\citep{louit2025calfsbm}. In addition, we
included methods using a single data source: $k$-means algorithms 
($k$-means)~\citep{lloyd1982least} applied only to covariates, and spectral 
clustering (SPC)~\citep{ng2002spectral} applied only to adjacency matrix. 
Since CASC, CASCORE, BCDC, CALF-SBM, and SPC each take one network as input,
we constructed an adjacency matrix $\bm A^\text{est}$ with entries 
$A^\text{est}_{ij}=\mathbb{I}(\sum_{m=1}^M \tilde{A}^{(m)}_{ij}\geq M/2)$, 
and used it as their network input. We gave the competing methods that cannot 
handle unknown $K$ (i.e., CASC, CASCORE, $k$-means and SPC) an additional 
advantage by assuming the knowledge of the true number of communities.

All methods have
implementations in~\proglang{R}~\citep{Rcore2026}. Specifically, $k$-means was implemented via
the~\pkg{stats} package, and SPC and CASC were performed using the~\pkg{c4}
package~\citep{c4package}. We used the~\pkg{CASCORE}
package~\citep{CASCOREpackage}, the~\pkg{bcdc}
package~\citep{bcdcpackage}, and the~\pkg{calfsbm}
package~\citep{calfsbmpackage} for the corresponding methods. For GBM, we
used the authors' publicly available code.

\subsection{Results}

\begin{figure}[tbp]
	\centering
	\includegraphics[width=0.95\textwidth]{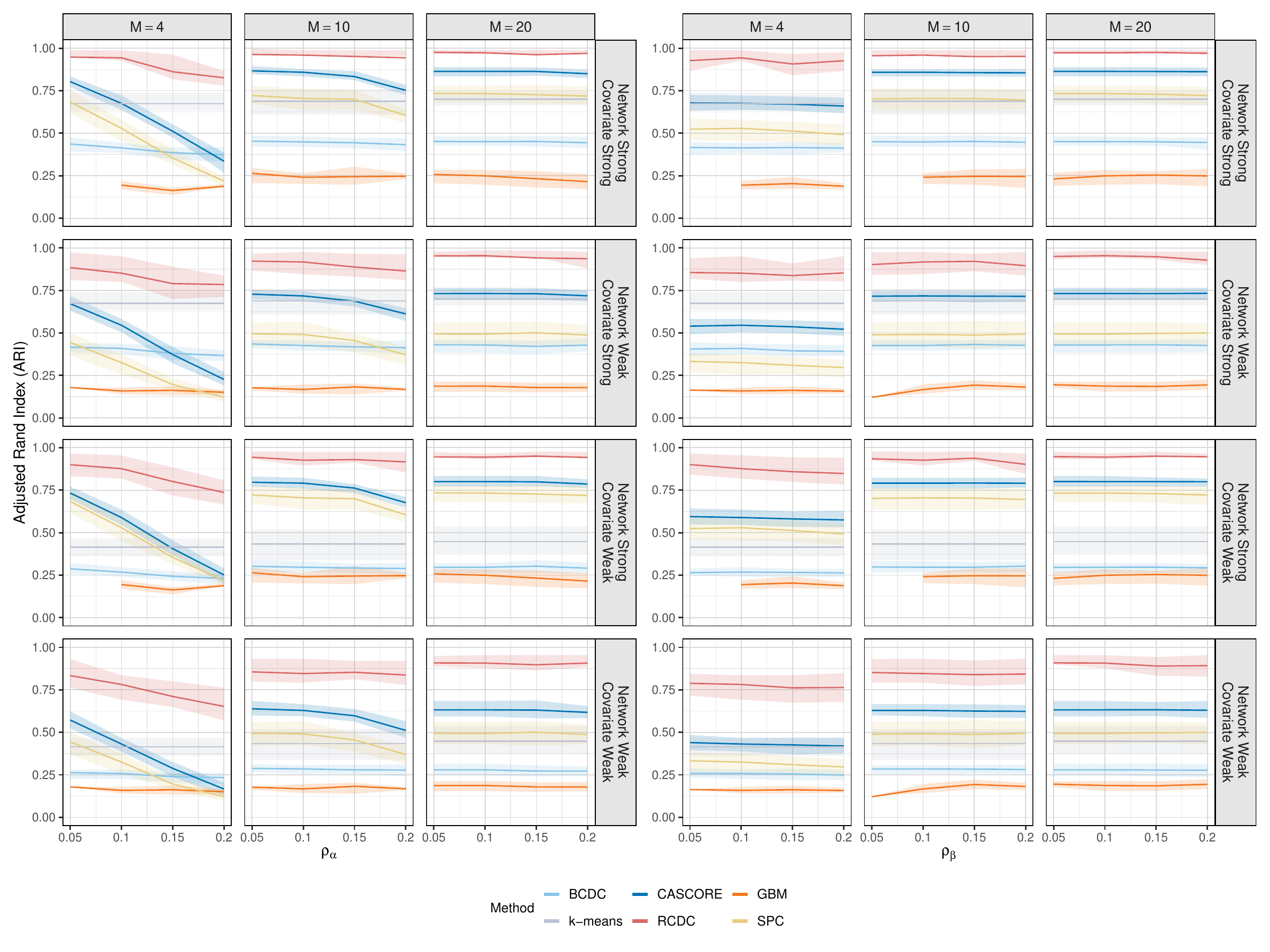}
	\caption{ARI results for different methods under homogeneous network
    noise ($\zeta_{\alpha}=\zeta_{\beta}=1$).
	 Left panel: fixing false negative rate at $0.1$ ($\rho_{\beta}=0.1$) and
     varying the other settings. 
	Right panel: fixing false positive rate at $0.1$ ($\rho_{\alpha}=0.1$)
    and varying the other settings.}
    \label{fig:fig1}
\end{figure}

Figure~\ref{fig:fig1} shows the mean ARI with the interquartile range for
RCDC and competing methods, averaged over 50 replicates, under homogeneous
network noise (i.e., $\zeta_{\alpha}=\zeta_{\beta}=1$).  
For clarity of presentation, we omit CALF-SBM, as it consistently performs
worse than or comparably to the other covariate-assisted community detection
method that accommodates unknown $K$ (i.e., BCDC). We also omit CASC, which
performs worse than or comparably to the other covariate-assisted method that
requires knowledge of $K$ (i.e., CASCORE) in almost all settings. Due to instability of 
the expectation–maximization algorithm in GBM, results could not be obtained 
for some parameter settings, and settings with fewer than 50 successful 
replicates are summarized using the mean of the available replicates. 
In nearly all other settings, GBM has the lowest ARI. 
RCDC outperforms all competing methods across all settings, including varying
network noise levels, numbers of noisy networks, and network structural and
covariate signal strengths. 

The left and right panels in Figure~\ref{fig:fig1} show the effect of
homogeneous false positive and false negative rates when fixing either false
positive rate or false negative rate at $0.1$ (i.e., $\rho_{\alpha}$ or
$\rho_{\beta}=0.1$) and varying the other settings.  
The ARIs of CASCORE and SPC decrease substantially as the false positive
rate increases when the number of noisy networks $M$ is $4$ (the left panel,
column 1), indicating sensitivity to false positive edges that weaken the
community structure. This effect becomes less pronounced as $M$ increases,
corresponding to less noisy input networks. When the false negative rate
increases (the right panel), the performance of all methods remains
relatively stable. One possible explanation is that increasing the false
negative rate removes both within- and between-community edges, and the true
networks are relatively sparse, thus the strength of the community structure
remains largely unchanged. Overall, our method demonstrates robust
performance across settings.

\begin{figure}[tbp]
	\centering
	\includegraphics[width=0.95\textwidth]{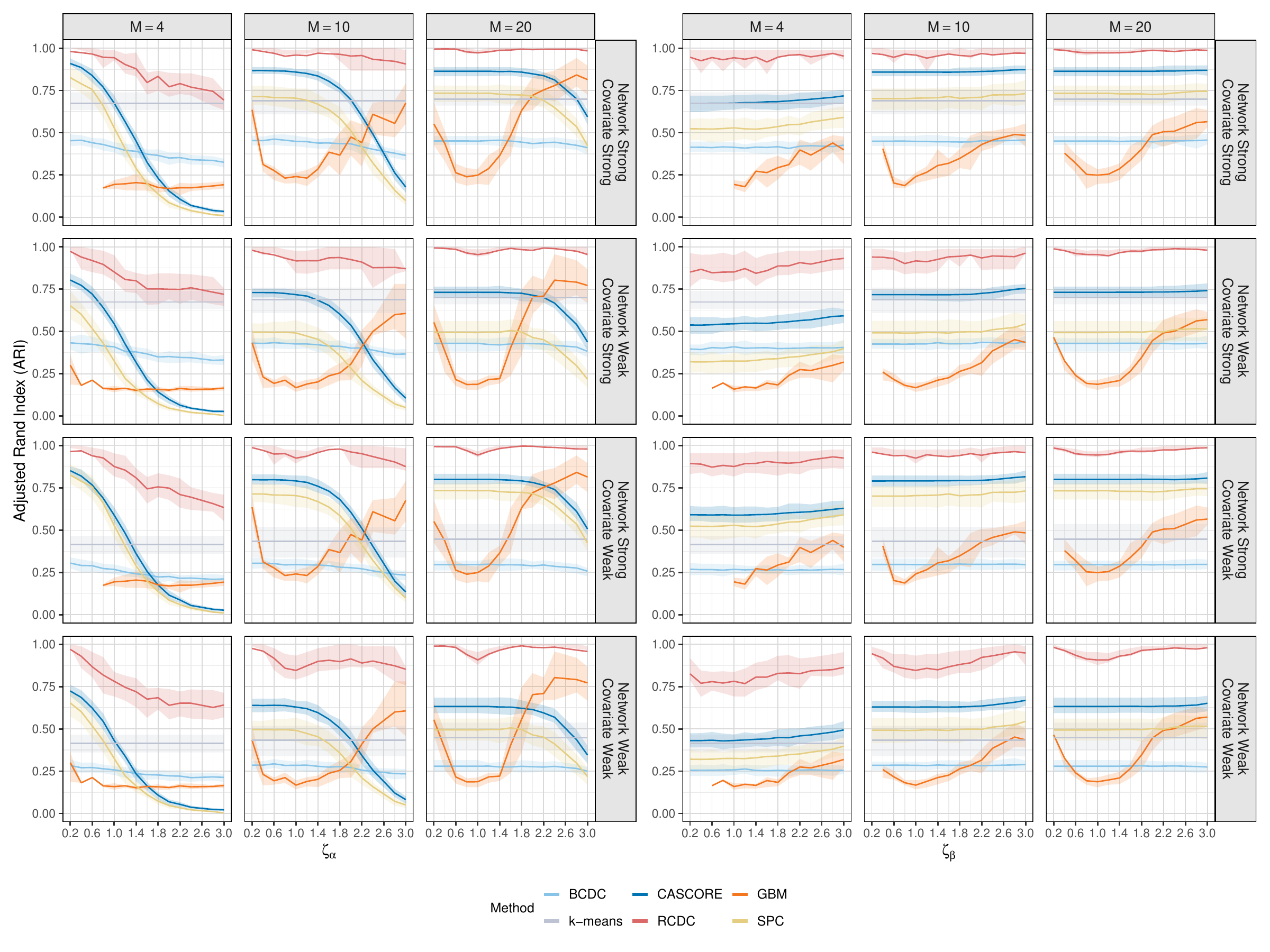}
	\caption{ARI results for different methods under heterogeneous network
    noise, with overall noise level parameters
    $\rho_{\alpha}=\rho_{\beta}=0.1$. 
	Left panel: fixing false negative rate at homogeneity ($\zeta_{\beta}=1$)
    and varying the other settings. 
	Right panel: fixing false positive rate at homogeneity
    ($\zeta_{\alpha}=1$) and varying the other settings.}
    \label{fig:fig2}
\end{figure}

The ARIs under heterogeneous network noise are presented in
Figure~\ref{fig:fig2}. We omit CALF-SBM and CASC again for clarity of
presentation, and some GBM results are unavailable due to algorithmic
instability. Across all settings, RCDC consistently achieves the highest ARI,
including scenarios with varying degrees of network noise heterogeneity,
different numbers of noisy networks, and varying strengths of structural and
covariate signals. 

The left and right panels in Figure~\ref{fig:fig2} show the effects of
heterogeneous false positive and false negative rates, respectively. We set
the overall noise level parameters $\rho_{\alpha}=\rho_{\beta}=0.1$, while
fixing either the false positive or false negative rate at homogeneity
($\zeta_{\alpha}$ or $\zeta_{\beta}=1$) and varying the other settings. When
$M=4$, the ARIs of CASCORE and SPC decrease substantially as the ratio of
between- to within-community false positive rates $\zeta_{\alpha}$ increases,
and increase when the ratio of between- to within-community false negative
rates $\zeta_{\beta}$ increases. This pattern implies that both methods are
particularly sensitive to between-community edges, which weaken the
underlying community structure. This pattern becomes less pronounced as $M$
increases, consistent with reduced noise in the input networks.

GBM shows a different form of sensitivity under heterogeneous network noise.
Although GBM explicitly models network noise, its performance is highly
sensitive to both the ratio of between- to within-community noise levels and
the number of noisy networks. Specifically, GBM performs poorly when $M=4$ and
displays a non-monotonic pattern when $M=10$ and $20$, with the lowest ARI
occurring around $\zeta_{\alpha},\zeta_{\beta}=1$ (i.e., homogeneous network
noise). This suggests that GBM requires a relatively large number of noisy
networks and sufficiently strong structural signal to effectively account for
network noise. The ARIs of RCDC are relatively similar across all settings,
which implies that its performance is robust to heterogeneity in
community-structured network noise.

In summary, these results suggest that leveraging both covariates and
noisy networks to infer the underlying community structure yields substantial
gains compared to existing approaches that either lack noise correction or
rely on a single data source.

\section{Data Application}
\label{sec:real_data}

We applied the proposed method to analyses of two 
functional brain network datasets, one from the ADNI 
study~\citep{mueller2005alzheimer} and the other
from the HCP~\citep{van2013wu}.

\subsection{Application to ADNI Data} 
\label{sec:ADNI}

We applied our method to ROI-level functional brain
networks with brain region covariate information constructed from data in the
ADNI database (\url{https://adni.loni.usc.edu}). The ADNI is a large, 
multi-center longitudinal study designed to investigate the progression of AD
using neuroimaging, biomarkers, and clinical
assessments~\citep{weiner2013alzheimer, weiner2017recent}. We used data from
the ADNI Phase~3 (ADNI-3) cohort. Resting-state functional 
magnetic resonance imaging (rs-fMRI) data were processed using the
ConnectomeMapper3 pipeline built in Nipype, and details of packages are
available in~\citet{tourbier2022} and related literature. The procedure for
obtaining blood--oxygen--level--dependent (BOLD) time signals of ROIs defined
by the Lausanne 2018 parcellation is described in \citet{xu2024topology}.
Covariate information, including cortical volume, surface area, and average
cortical thickness for 68 Desikan–Killiany regions, was obtained from the UC
Berkeley processing pipeline~\citep{landau2015measurement} and matched to
imaging scan dates.

For each subject, we constructed a functional brain 
network for the 68 regions based on the Fisher $z$-transformed Pearson
correlation between BOLD time signals of pairs of ROIs, thresholded at
$0.4$~\citep{garrison2015stability, wang2014systematic}. We excluded subjects
with missing BOLD time series or diagnostic status. Our analysis included 391
subjects, consisting of 188 healthy control (HC), 150 mild cognitive
impairment (MCI), and 53 dementia (AD). The demographic characteristics are
summarized in Table~\ref{tab:demographics}. We applied our model to detect 
communities for each diagnostic group.

\begin{table}[tbp]
	\centering
	\begin{threeparttable}
	\caption{Demographic characteristics of subjects in the ADNI dataset by
    diagnosis group.}
    \label{tab:demographics}
	\begin{tabular}{lccc}
        \toprule
        Characteristic & HC (n=188) & MCI (n=150) & AD (n=53) \\
        \midrule
        Age, years (mean $\pm$ SD) & 70.6 $\pm$ 6.6 & 71.7 $\pm$ 7.1 & 74.7 $\pm$ 7.0 \\
        \\
        Gender, n (\%) & & & \\
        \quad Female & 98 (52.1) & 68 (45.3) & 23 (43.4) \\
        \quad Male & 90 (47.9) & 82 (54.7) & 30 (56.6) \\
        \\
        Race, n (\%) & & & \\
        \quad White & 168 (89.4) & 142 (94.7) & 51 (96.2) \\
        \quad Black & 13 (6.9) & 3 (2.0) & 0 (0.0) \\
        \quad Asian & 1 (0.5) & 2 (1.3) & 1 (1.9) \\
        \quad Indian/Alaskan & 1 (0.5) & 0 (0.0) & 0 (0.0) \\
        \quad More than one & 4 (2.1) & 2 (1.3) & 1 (1.9) \\
        \quad Unknown & 1 (0.5) & 1 (0.7) & 0 (0.0) \\
        \\
        Ethnicity, n (\%) & & & \\
        \quad Hispanic/Latino & 11 (5.9) & 5 (3.3) & 1 (1.9) \\
        \quad Not Hispanic/Latino & 174 (92.6) & 145 (96.7) & 52 (98.1) \\
        \quad Unknown & 3 (1.6) & 0 (0.0) & 0 (0.0) \\
        \\
        APOE4 status, n (\%) & & & \\
        \quad Non-carrier & 120 (63.8) & 85 (56.7) & 21 (39.6) \\
        \quad Carrier & 66 (35.1) & 58 (38.7) & 29 (54.7) \\
        \quad Unknown & 2 (1.1) & 7 (4.7) & 3 (5.7) \\
        \bottomrule
	\end{tabular}
	\begin{tablenotes}[flushleft]
        \footnotesize
        \item Note: SD denotes standard deviation.
    \end{tablenotes}
	\end{threeparttable}
\end{table}

Clustering results were first compared to 7 functional Resting-State 
Networks (RSNs) in the Yeo Atlas~\citep{yeo2011organization}, which were
derived from rs-fMRI data collected from healthy young adults.
Agreement with RSNs was highest in the HC group (0.340), and decreased 
progressively in the MCI (0.209) and AD (0.150) groups, 
indicating increasing disruption of functional community structure 
with disease progression. To further examine differences in community 
structure across diagnostic groups, we quantified the similarity of 
MCI and AD to HC using ARI. Similarity also decreased with
disease severity, with higher similarity in MCI (0.577) than in AD (0.506),
indicating a gradual reorganization of functional community structure 
from HC to MCI to AD.

\begin{figure}[tbp]
  \centering

  \begin{subfigure}[tbp]{\textwidth}
    \centering
    \caption{Community assignments for ROIs (R: right 
    hemisphere; L: left hemisphere).}
    \includegraphics[width=\textwidth]{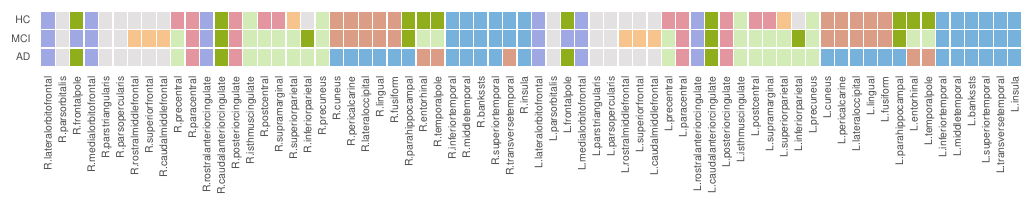}
    \label{fig:assignments}
  \end{subfigure}

  \begin{subfigure}[tbp]{\textwidth}
    \centering
    \caption{Spatial depiction of detected communities.}
    \includegraphics[width=\textwidth]{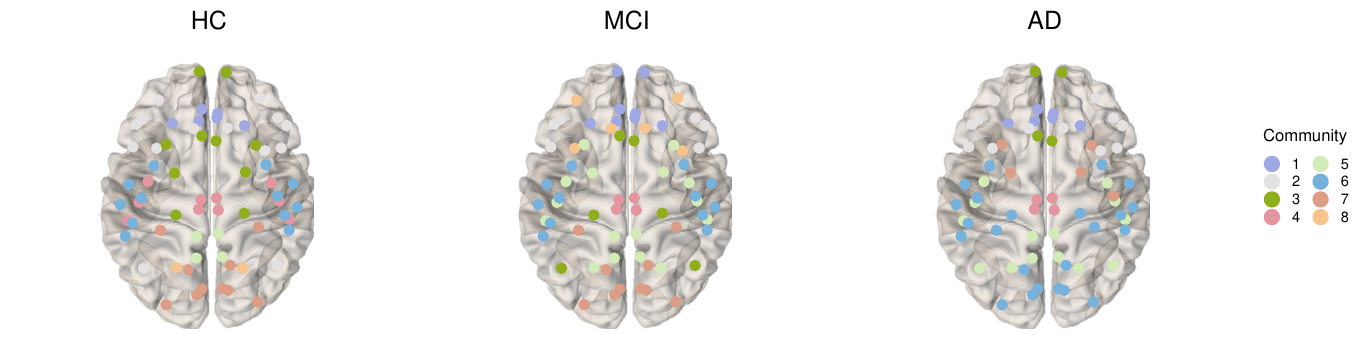}
    \label{fig:spatial}
  \end{subfigure}
  \caption{Community structure of functional brain networks in HC, MCI, and AD groups.}
  \label{fig:comm_structure}
\end{figure}

We also examined region-specific community assignments to identify 
localized patterns of reorganization (Figure~\ref{fig:comm_structure}). 
The number of detected communities was $K=8$ for HC and MCI and $K=7$ for AD.
Community labels for MCI and AD were aligned with those of HC using the
Hungarian algorithm~\citep{Kuhn1955} to maximize overlap for comparison.
Figure~\ref{fig:assignments} shows community assignments 
for ROIs in both hemispheres for HC, MCI, and AD groups,
where each region is color-coded according to its community membership.
Figure~\ref{fig:spatial} visualizes the consensus communities in 3D Montreal
Neurological Institute space, with nodes overlaid on a brain template to
illustrate functional topography across the three diagnostic groups.

Several ROIs known to be vulnerable in AD showed consistent 
and progressive shifts in community affiliation across disease 
stages (Figure~\ref{fig:assignments}). In particular, the entorhinal 
cortex, a key medial temporal
structure for memory processing, exhibited clear stage dependent changes in
community affiliation (from community 3 to 5, then 7), 
consistent with its role as an early site of 
neurodegeneration~\citep{braak1991neuropathological, killiany2000use}.
In addition, the inferior parietal cortex, a central hub of the 
default mode network, also exhibited consistent transitions across 
stages (from community 2 to 3, then 5), reflecting disruption of 
functional networks that are characteristically affected in 
AD~\citep{buckner2005molecular, greicius2004default}. In contrast,
prefrontal regions, including the rostral middle frontal, superior frontal, 
and caudal middle frontal cortices, exhibited stage-specific changes, 
forming a distinct community in the MCI stage. Such reorganization in 
prefrontal networks has been reported in MCI and is often interpreted as 
potential compensatory recruitment in response to early brain 
disruption~\citep{grady2003evidence, qi2010impairment}.

Finally, Figure~\ref{fig:spatial} shows that, in the AD group, community
boundaries became markedly less distinct, accompanied by the emergence of a
large, widely distributed community (blue) that incorporated many regions
assigned to communities 6 and 7 in HC and MCI groups. This pattern indicates
reduced network segregation and increased global integration, consistent with
widespread disruption of functional organization in advanced stages of
AD~\citep{brier2014functional}.

\subsection{Application to HCP Data}
\label{sec:HCP}

We applied our method to ROI-level functional brain networks 
with node-level covariate information constructed from the HCP data. The HCP 
is a large-scale project designed to characterize the structural and 
functional connectivity of the human brain~\citep{van2013wu}. 
We used rs-fMRI data from the young adult data release of the HCP, with four
runs acquired per subject. BOLD time series were obtained for 360 ROIs
defined by the HCP multimodal parcellation atlas~\citep{glasser2016multi},
following the procedure described in \citet{akiki2019determining}.
We considered ROI-level covariate information, including cortical thickness,
cortical curvature, and sulcal depth.

We constructed one functional brain network per subject per run using the
same Fisher z-transformed Pearson correlation approach described in
Section~\ref{sec:ADNI}. We selected thresholds to yield network densities
around 0.15~\citep{garrison2015stability}.
For each subject, we treated the four functional brain networks as noisy
realizations of the underlying network. Our analysis included 376 subjects
with complete BOLD time series and covariate information. The demographic
characteristics are summarized in Table~\ref{tab:hcp_demographics_sex}.
We applied our method to detect communities for each subject.

\begin{table}[tbp]
	\centering
	\begin{threeparttable}
	\caption{The demographic characteristics of subjects in the HCP dataset.}
	\label{tab:hcp_demographics_sex}
	\begin{tabular}{lcc}
        \toprule
        Characteristic & Female (n=198) & Male (n=178) \\
        \midrule
        Age, years (mean $\pm$ SD) & 29.4 $\pm$ 3.6 & 27.9 $\pm$ 3.7 \\
        \\
        BMI, mean $\pm$ SD & 25.76 $\pm$ 5.01 & 26.64 $\pm$ 4.27 \\
        \\
        Race, n (\%) & & \\
        \quad White & 146 (73.7) & 140 (78.7) \\
        \quad Black or African Am. & 33 (16.7) & 21 (11.8) \\
        \quad Asian/Nat. Hawaiian/Othr Pacific Is. & 10 (5.1) & 13 (7.3) \\
        \quad More than one & 3 (1.5) & 2 (1.1) \\
        \quad Unknown or Not Reported & 6 (3.0) & 2 (1.1) \\
        \\
        Ethnicity, n (\%) & & \\
        \quad Not Hispanic/Latino & 180 (90.9) & 157 (88.2) \\
        \quad Hispanic/Latino & 15 (7.6) & 19 (10.7) \\
        \quad Unknown or Not Reported & 3 (1.5) & 2 (1.1) \\
        \bottomrule
	\end{tabular}
	\end{threeparttable}
\end{table}

\begin{figure}[tbp]
	\centering
	\includegraphics[width=0.95\textwidth]{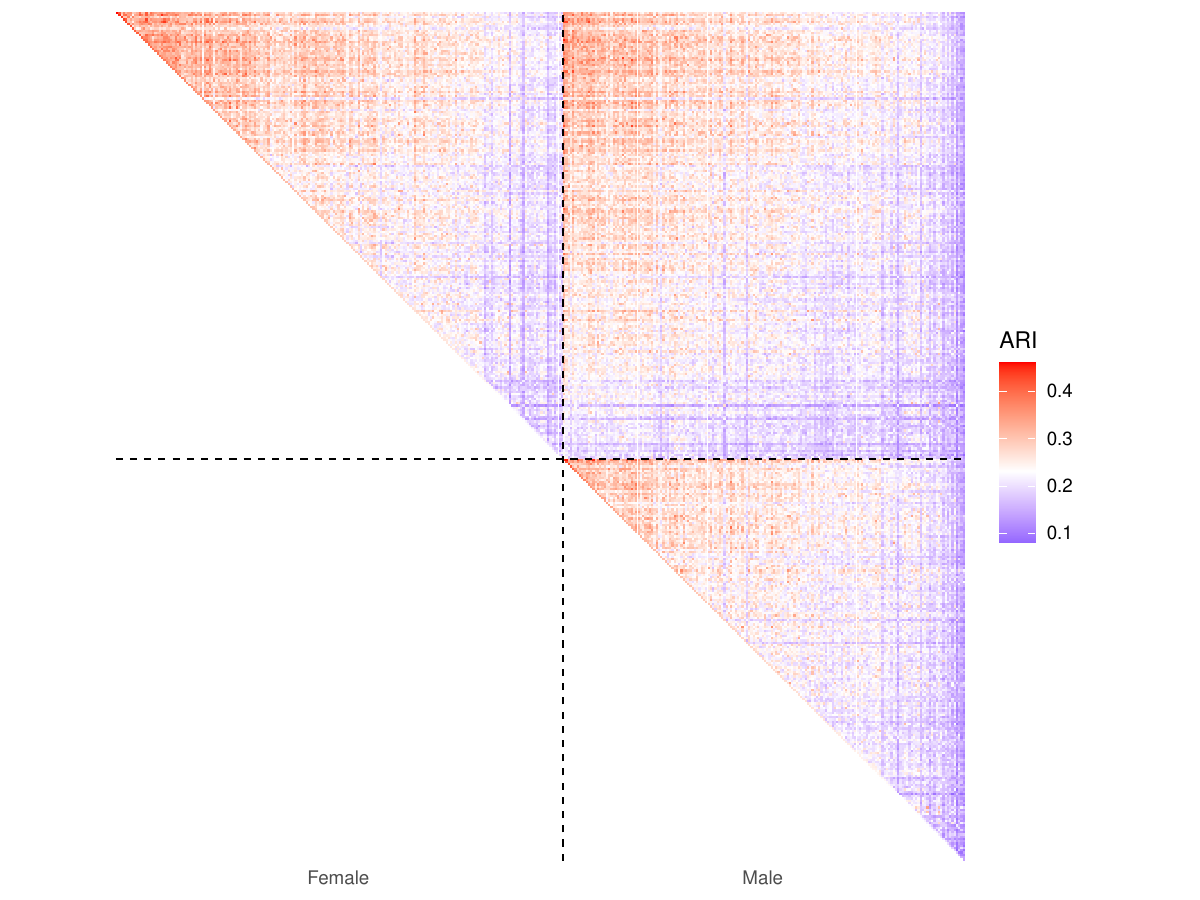}
	\caption{Heat map of ARIs for the community partitions of 376 subject-level 
    brain networks. Each row and column corresponds to a subject. Off-diagonal 
    entries show pairwise ARIs between subjects, while diagonal 
    entries show the ARI between each subject-specific community assignment 
    and CAB-NP. Cells are colored on a diverging blue--white--red scale, with 
    blue and red denoting ARI values below and above the median pairwise ARI, 
    respectively, and white corresponding to the median. Dashed lines separate 
    female and male subjects.}
    \label{fig:hcp_heatmap}
\end{figure}

Figure~\ref{fig:hcp_heatmap} displays the ARI heat map. Off-diagonal entries
show pairwise ARIs between subjects, and diagonal entries show the ARI
between each subject-specific community assignment and Cole--Anticevic Brain-wide 
Network Partition~\citep[CAB-NP,][]{ji2019mapping}, a partition of the ROIs
into 12 canonical functional networks. The subject-specific communities
showed moderate agreement with CAB-NP, and the pairwise ARIs were positive
and varied across subjects. These implied that our method detects community
structures that capture shared functional organization and preserve subject-level
heterogeneity. Females showed higher ARI than males for CAB-NP, 
and pairwise ARIs were higher among female subjects than among male subjects. 
These findings are consistent with previous studies reporting differences
between females and males in functional brain
networks~\citep{filippi2013organization,zhang2018functional}.

\section{Discussion}
\label{sec:disc}

We have developed a Bayesian framework for detecting the underlying
community structure from noisy networks with node-level covariates. By
jointly leveraging information from noisy graphs and node covariates, we
provide a robust approach to recovering the community structure of the
true underlying network in the presence of network noise. Simulation studies
show improved performance of the proposed method over several existing
approaches. Analyses of functional brain networks from the HCP and 
ADNI datasets show that our method yields interpretable community structures, 
captures subject-specific variation consistent with 
population-level functional organization in healthy young adults, and 
identifies systematic patterns of network reorganization associated with 
disease severity.

Beyond neuroscience, our method is applicable to a wide range of fields
involving noisy networks and node-level covariates. For example, in social
networks, edges representing interactions or friendships are usually
partially observed or misreported~\citep{butts2003network, 
kossinets2006effects, newman2018network}, while node covariates may include
demographic and behavioral attributes. In biological networks, protein--protein 
or gene interaction networks are often contaminated by
experimental false positives and negatives~\citep{chen2006increasing, 
hart2006complete, sprinzak2003reliable, von2002comparative}, while node covariates may
include gene expression profiles and functional annotations. Jointly
leveraging noisy network information and node-level covariates is
important for recovering interpretable latent communities across diverse
application domains.

Our work also suggests several extensions. First, extending the
framework to weighted noisy networks would allow for the use of additional
information to better infer the underlying community structure. Second, it
would be useful to consider network noise with
dependence and/or unstructured heterogeneity. Finally, using
nonparametric Bayesian priors to infer the number of communities and
developing alternative algorithms such as variational expectation-maximization
could reduce the computational cost.

\appendix
\section{Appendix: MCMC Algorithm}
\label{sec:app}

In this section, we develop an efficient MCMC algorithm for sampling the 
parameters in our model. We first introduce a data augmentation technique 
used to facilitate the sampling. We then present the conditional posterior 
distributions of the model parameters. Finally, we describe the 
initialization procedure.

\subsection*{A.1 Data Augmentation}
\label{sec:data_aug}

We introduce P$\acute{\text{o}}$lya-Gamma variables for $\bm A$. 
For simplicity, we denote
\begin{align*}
	\psi_{ij}=B_{z_i z_j}\mathbb{I}(z_i\neq z_j)+\theta_i+\theta_j.
\end{align*}
We note that, for $\ 1\leq i<j\leq n $, 
\begin{align}
	p(A_{ij}\mid z_i,z_j,B_{z_i z_j},\theta_i,\theta_j)
	=\frac{\exp(\psi_{ij})^{A_{ij}}}{1+\exp(\psi_{ij})}.
	\label{bernoulli_logit}
\end{align}
Using Theorem 1 in \cite{polson2013bayesian}, we can introduce an 
edge-specific latent P$\acute{\text{o}}$lya-Gamma variable 
$w_{ij}$ ($\mathcal{PG}(1,0)$) and rewrite the right hand side 
of~\eqref{bernoulli_logit} as:
\begin{align*}
	\frac{\exp(\psi_{ij})^{A_{ij}}}{1+\exp(\psi_{ij})}
	= \frac{1}{2}\exp(\kappa_{ij}\psi_{ij})
	\int_0^\infty \exp\left(-\frac{1}{2}w_{ij}\psi_{ij}^2\right)
	p(w_{ij})\,dw_{ij},
\end{align*}
where $\kappa_{ij}=A_{ij}-1/2$. This leads to the conditional distribution 
of $A_{ij}$
\begin{align*}
	p(A_{ij}\mid z_i,z_j,B_{z_i z_j},\theta_i,\theta_j,w_{ij})
	\propto \exp\!\left(\kappa_{ij}\psi_{ij}-\tfrac{1}{2}w_{ij}
	\psi_{ij}^2\right).
\end{align*}

\subsection*{A.2 Conditional Posterior Distributions}
\label{sec:post}
 
We generate samples from the posterior density. Let $\bm{X} = 
(\bm{x}_1,\dots,\bm{x}_n)^\top$ denote the matrix of 
$p$-dimensional nodal covariates, including $R$ continuous 
and $S$ categorical variables, and let $\tilde{\bm{A}}^{(1:M)} = 
\{\tilde{\bm{A}}^{(m)}: m=1,\dots,M\}$ be the collection of all $M$
noisy adjacency matrices. Let $\Theta$ denote all
model parameters. The posterior density of $\Theta$ is given by, 
\begin{align*}
    p(\Theta|\bm X, \tilde{\bm A}^{(1:M)} ) 
    &\propto p(\bm{X} \mid \bm{z}, \bm{\mu}, \bm{\tau}^2, \bm{\xi}) \; 
     p(\bm{\mu} \mid \bm{\tau}^2) \; p(\bm{\tau}^2)
    \; p(\bm{\xi}) \; p(\bm{z} \mid \bm{\pi}) \; p(\bm{\pi})
    \\
    &\quad \times
    p(\tilde{\bm{A}}^{(1:M)} \mid \bm{A}, \bm{z}, \bm{\alpha}, \bm{\beta})
    \; p(\bm{\alpha}) \; p(\bm{\beta})\;  p(\bm{A} \mid \bm{z},
    \bm{\theta}, \bm{B},\bm w) \; p(\bm{\theta}) \; p(\bm{B}) \; p(\bm{w}) ,
\end{align*}
where $\bm w = \{w_{ij}:1\leq i<j\leq n\}$,  
$\bm{\alpha} = (\alpha_{kl})_{K\times K}$, $\bm{\beta} = 
(\beta_{kl})_{K\times K}$; $\bm{\mu} = (\mu_{kr})_{K\times R}$,
$\bm{\tau}^2 = (\tau^2_{kr})_{K\times R}$,  and  
$\bm{\xi} = \{\bm{\xi}_{ks}:k=1,\dots,K, s=1,\dots,S\}$.
 
Next, we show the conditional distribution for each parameter. 

\textit{(1) The conditional distribution of $A_{ij}$.}
We use Gibbs sampling to sample $A_{ij}$ from its conditional 
distribution, that is 
\begin{align*}
    A_{ij} \mid \tilde{\bm{A}}^{(1:M)}, z_i, z_j, B_{z_iz_j}, 
    \theta_i, \theta_j, \alpha_{z_iz_j}, \beta_{z_iz_j} \sim 
    \mathrm{Bern}(p_{ij}^*),
\end{align*}
where
\begin{align*}
p_{ij}^* = \text{logit}^{-1} \Bigg(
  \psi_{ij} + \ln \Bigg[
    \Big(\tfrac{1-\beta_{z_iz_j}}{\alpha_{z_iz_j}}\Big)^
    {\sum_{m=1}^M \mathbb{I}(\tilde A_{ij}^{(m)} = 1)}
    \Big(\tfrac{\beta_{z_iz_j}}{1-\alpha_{z_iz_j}}\Big)^
    {\sum_{m=1}^M \mathbb{I}(\tilde A_{ij}^{(m)} = 0)}
  \Bigg]
\Bigg).
\end{align*}

\textit{(2) The conditional distribution of $w_{ij}$.}
We use Gibbs sampling to sample $w_{ij}$ from its conditional 
distribution, that is
\begin{align*}
    w_{ij} \mid A_{ij}, z_i, z_j, B_{z_iz_j}, \theta_i, \theta_j 
    \sim \mathcal{PG}(1, \psi_{ij}).
\end{align*}

\textit{(3) The conditional distributions of $\bm B$ and $\bm \theta$.} 
We define $\bm{\delta} = (\bm{b}^{\top},
\bm{\theta}^{\top})^{\top}$, 
where $\bm{b} = (\{B_{kl} : 1 \le k < l \le K\})^{\top}$, with
the pairs $(k,l)$ ordered column-wise over the upper triangle,
i.e., $(1,2),(1,3),(2,3),\ldots,(K-1,K)$. 
Let $\bm{Y}$ denote the design matrix associated with Model (2),
of dimension $\binom{n}{2} \times \left[\binom{K}{2} + n\right]$.
Its rows are indexed by the node pairs $(i,j)$ with $i<j$,
ordered column-wise over the upper triangle, and its columns
correspond to the entries of $\bm{\delta}$. Let
$\bm{y}_{ij}^{\top}$ denote the row of $\bm{Y}$ corresponding to
the node pair $(i,j)$. Then, we have
\begin{align*}
    \bm{y}_{ij}^{\top} \bm{\delta}
    = B_{z_i z_j}\mathbb{I}(z_i \neq z_j) + \theta_i + \theta_j
    = \psi_{ij}.
\end{align*}

As an example, consider $n=4$ nodes with $K=3$ communities and
$\bm{z} = (1,1,2,3)$. In this case
$\bm{\delta} = (B_{12}, B_{13}, B_{23},
\theta_1, \theta_2, \theta_3, \theta_4)^{\top}$, and the design matrix
$\bm{Y}$ is
\begin{align*}
\bm{Y} =
\begin{blockarray}{ccc|cccc l}
B_{12} & B_{13} & B_{23} & \theta_1 & \theta_2 & \theta_3 & \theta_4 & \\
\begin{block}{[ccc|cccc]l}
  0 & 0 & 0 & 1 & 1 & 0 & 0 & \ (1,2) \\
  1 & 0 & 0 & 1 & 0 & 1 & 0 & \ (1,3) \\
  1 & 0 & 0 & 0 & 1 & 1 & 0 & \ (2,3) \\
  0 & 1 & 0 & 1 & 0 & 0 & 1 & \ (1,4) \\
  0 & 1 & 0 & 0 & 1 & 0 & 1 & \ (2,4) \\
  0 & 0 & 1 & 0 & 0 & 1 & 1 & \ (3,4) \\
\end{block}
\end{blockarray}.
\end{align*}

The conditional distribution of $\bm{\delta}$ is 
\begin{align*}
    \bm{\delta} \mid \bm{A}, \bm{w}, \bm{z} 
    &\sim \mathcal{N}(\bm{\eta}, \bm{V}) \, \mathbb{I}(\bm{b} \le \bm{0}),
\end{align*}

where $\bm{V} = \left( \bm{Y}^\top \bm{\Omega} \bm{Y} + \frac{1}{\sigma^2} 
\bm{I}_{n + K(K-1)/2} \right)^{-1}$ and 
$\bm{\eta} = \bm{V} \bm{Y}^\top \bm{\kappa}$, with 
$\bm{\kappa} = \{\kappa_{ij} :1 \le i < j \le n\}^{\top}$ and
$\bm{\Omega} = \text{diag}(\{w_{ij}:1 \le i < j \le n\})$,
ordered consistently with the rows of $\bm Y$.

We partition the parameters, mean, and covariance matrix into blocks
corresponding to $\bm b$ and $\bm{\theta}$:
\begin{align*}
    \bm{\delta} = \begin{pmatrix} \bm{b} \\ 
    \bm{\theta} \end{pmatrix}, \quad
    \bm{\eta} = \begin{pmatrix} \bm{\eta}_b \\ 
    \bm{\eta}_\theta \end{pmatrix}, \quad
    \bm{V} = \begin{pmatrix} \bm{V}_{bb} & \bm{V}_{b\theta} \\ 
    \bm{V}_{\theta b} & \bm{V}_{\theta\theta} \end{pmatrix}.
\end{align*}
We sample $\bm{\delta}$ by first sampling 
$\bm{\theta}$ marginally from its conditional distribution, 
\begin{align*}
	\bm{\theta} \mid \bm{A}, \bm{w}, \bm{z} &\sim 
    \mathcal{N}(\bm{\eta}_\theta, \bm{V}_{\theta\theta}),
\end{align*}
followed by sampling $\bm b$ from a truncated normal distribution
\begin{align*}
    \bm{b} \mid \bm{\theta}, \bm{A}, \bm{w}, \bm{z} &\sim 
    \mathcal{N} \left( \bm{\eta}_b + \bm{V}_{b\theta} 
    \bm{V}_{\theta\theta}^{-1} (\bm{\theta} - \bm{\eta}_\theta), 
    \bm{V}_{bb} - \bm{V}_{b\theta} \bm{V}_{\theta\theta}^{-1} 
    \bm{V}_{\theta b} \right) \mathbb{I}(\bm{b} \le \bm{0}),
\end{align*}
and then reconstructing the matrix $\bm{B}$ from $\bm{b}$.

\textit{(4) The conditional distributions of $\alpha_{kl}$ and 
$\beta_{kl}$.} We use Gibbs sampling to sample $\alpha_{kl}$ and 
$\beta_{kl}$ from their conditional distributions, which are 
\begin{align*}
	\alpha_{kl} \mid \tilde{\bm{A}}^{(1:M)}, \bm{A}, \bm{z} 
	&\sim \text{Beta}\biggl( 
	a_{\alpha} + \sum_{1 \le i < j \le n} 
	\mathbb{I}\bigl(\{z_i,z_j\}=\{k,l\}\bigr) 
	\mathbb{I}(A_{ij}=0) 
	\sum_{m=1}^{M} \mathbb{I}(\tilde{A}_{ij}^{(m)}=1), \\
	&\qquad b_{\alpha} + \sum_{1 \le i < j \le n} 
	\mathbb{I}\bigl(\{z_i,z_j\}=\{k,l\}\bigr)
	\mathbb{I}(A_{ij}=0) 
	\sum_{m=1}^{M} \mathbb{I}(\tilde{A}_{ij}^{(m)}=0) 
	\biggr), \\
	\beta_{kl} \mid \tilde{\bm{A}}^{(1:M)}, \bm{A}, \bm{z} 
	&\sim \text{Beta}\biggl( 
	a_{\beta} + \sum_{1 \le i < j \le n} 
	\mathbb{I}\bigl(\{z_i,z_j\}=\{k,l\}\bigr)
	\mathbb{I}(A_{ij}=1) 
	\sum_{m=1}^{M} \mathbb{I}(\tilde{A}_{ij}^{(m)}=0), \\
	&\qquad b_{\beta} + \sum_{1 \le i < j \le n} 
	\mathbb{I}\bigl(\{z_i,z_j\}=\{k,l\}\bigr)
	\mathbb{I}(A_{ij}=1) 
	\sum_{m=1}^{M} \mathbb{I}(\tilde{A}_{ij}^{(m)}=1) 
	\biggr).
\end{align*}

\textit{(5) The conditional distributions of continuous covariate 
parameters $\mu_{k r}$ and $\tau^2_{k r}$.} We use Gibbs sampling to 
sample $\mu_{k r}$ and $\tau^2_{k r}$ from their conditional 
distributions, which are
\begin{align*}
	\mu_{k r} \mid \{x_{i r}^{\text{cont}}:z_i=k\}, \tau^2_{k r} 
	&\sim \mathcal{N}\left( 
	\frac{\mu_{0 r}/\iota_{0 r} + \sum_{i=1}^n \mathbb{I}(z_i=k)\, 
	x_{i r}^{\text{cont}}}{1/\iota_{0 r} + N_k}, \quad
	\frac{\tau^2_{k r}}{1/\iota_{0 r} + N_k} 
	\right), \\
	\tau^2_{kr} \mid \{x_{ir}^{\text{cont}}:z_i=k\}, \mu_{kr}
&\sim \mathcal{IG}\Bigg(
a_{\tau r} + \frac{1 + N_k}{2}, \\
&\qquad b_{\tau r} + \frac{(\mu_{kr} - \mu_{0r})^2}{2\iota_{0r}}
+ \frac{1}{2} \sum_{i=1}^n \mathbb{I}(z_i = k)
\left(x_{ir}^{\text{cont}} - \mu_{kr}\right)^2
\Bigg).
\end{align*}

\textit{(6) The conditional distribution of categorical covariate 
parameters $\bm{\xi}_{ks}$.} We use Gibbs sampling to sample 
$\bm{\xi}_{ks}$ from its conditional distribution, that is 
\begin{align*}
	\bm{\xi}_{ks} \mid \{x_{i s}^{\text{cat}}:z_i=k\}
	&\sim \text{Dir} \left( 
	\gamma_{s} + \sum_{z_i=k} \mathbb{I}(x_{i s}^{\text{cat}}=1), \;
	\dots, \;
	\gamma_{s} + \sum_{z_i=k} \mathbb{I}(x_{i s}^{\text{cat}}=C_s) 
	\right).
\end{align*}

\textit{(7) The conditional distributions of community mixing 
proportions $\bm{\pi}$.} We use Gibbs sampling to sample $\bm{\pi}$ from
its conditional distribution, that is 
\begin{align*}
    \bm{\pi} \mid \bm{z}, d &\sim \text{Dir}\left( \frac{d}{K} + 
    N_1, \dots, \frac{d}{K} + N_K \right).
\end{align*}

\textit{(8) The conditional distribution of community label $\bm z$.} We 
use Gibbs sampling to sample~$\bm z$ from its conditional distribution, 
that is
\begin{align*}
& p(z_i = k \mid \tilde{\bm{A}}^{(1:M)}, \bm{A}, \bm{w}, 
\bm{X}, \bm{z}_{-i}, \bm{\pi}, 
\bm{\theta}, \bm{B}, \bm{\mu}, \bm{\tau}^2, \bm{\xi}, \bm{\alpha}, 
\bm{\beta}) \\
&\propto \pi_k \times \prod_{j \neq i} \prod_{m=1}^{M} 
\left[
(1-\beta_{k z_j})^{\mathbb{I}(\tilde{A}_{ij}^{(m)}=1)} 
\beta_{k z_j}^{\mathbb{I}(\tilde{A}_{ij}^{(m)}=0)}
\right]^{A_{ij}} 
\left[
\alpha_{k z_j}^{\mathbb{I}(\tilde{A}_{ij}^{(m)}=1)} 
(1-\alpha_{k z_j})^{\mathbb{I}(\tilde{A}_{ij}^{(m)}=0)}
\right]^{1-A_{ij}} \\
&\quad \times 
\exp\left(
\sum_{j \neq i} 
\left[
\kappa_{ij} \psi_{ij}^{(k)} 
- \frac{1}{2} w_{ij} (\psi_{ij}^{(k)})^2
\right]
\right) \\
&\quad \times \prod_{r=1}^R 
\frac{1}{\sqrt{\tau^2_{k r}}}
\exp\left(
-\frac{(x_{i r}^{\text{cont}} - \mu_{k r})^2}{2\tau^2_{k r}}
\right) \\
&\quad \times \prod_{s=1}^S 
\xi_{k s x_{i s}^{\text{cat}}},
\end{align*}
where $\psi_{ij}^{(k)} = B_{k z_j} \mathbb{I}(k \neq z_j) + \theta_i 
+ \theta_j$, $\bm{z}_{-i} = (z_1,\dots,z_{i-1},z_{i+1},\dots,z_n)$ denotes 
the community assignments of all nodes except node $i$.

\subsection*{A.3 Initialization}
\label{sec:init}

Motivated by \citet{le2018estimating}, we propose the following 
initialization strategy to facilitate faster MCMC convergence. 
 
\textit{(1) Initialization of $\bm A$}: We set
$A^{\text{init}}_{ij}=\mathbb{I}(\sum_{m=1}^M \tilde{A}^{(m)}_{ij} \ge M/2)$.

\textit{(2) Initialization of $\bm z$}: We initialize the community
assignments $\bm z^{\text{init}}$ using spectral
clustering~\citep{ng2002spectral} applied to the initial adjacency matrix
$\bm A^{\text{init}}$.

\textit{(3) Initialization of $\bm \theta$ and $\bm B$}: We first 
calculate the edge density between community $k$ and $l$ based on $\bm
A^{\text{init}}$ and $\bm z^{\text{init}}$, which is given by
\begin{align*}
	o^{\text{init}}_{kl} &= \frac{ \sum_{i<j} 
	\mathbb{I}\bigl(\{z_i^{\text{init}},z_j^{\text{init}}\}=\{k,l\}\bigr)
	A^{\text{init}}_{ij}} {\sum_{i<j} 
	\mathbb{I}\bigl(\{z_i^{\text{init}},z_j^{\text{init}}\}=\{k,l\}\bigr)}.
\end{align*}
We initialize $\theta_i^{\text{init}} = \frac{1}
{2}\,\mathrm{logit}\!\left(o^{\text{init}}_{kk}\right)$ for all $i\in
\{i:z_i^{\text{init}}=k\}$, and set $B_{kl}^{\text{init}} = 
\mathrm{logit}\!\left(o^{\text{init}}_{kl}\right) - \frac{1}
{2}\,\mathrm{logit}\!\left(o^{\text{init}}_{kk}\right) - \frac{1}
{2}\,\mathrm{logit}\!\left(o^{\text{init}}_{ll}\right)$.

\textit{(4) Initialization of $\alpha_{kl}$ and $\beta_{kl}$}: We 
initialize $\alpha_{kl}$ and $\beta_{kl}$ based on the false positive and
false negative rates of noisy networks relative to $\bm A^{\text{init}}$.
Specifically, we set
\begin{align*}
	\alpha_{kl}^{\text{init}} &= \frac{ \sum_{i<j}
	\mathbb{I}\bigl(\{z_i^{\text{init}},z_j^{\text{init}}\}=\{k,l\}\bigr)\, 
	\Bigl(\sum_{m=1}^M \tilde{A}^{(m)}_{ij}\Bigr)\, 
	\mathbb{I}\bigl(A^{\text{init}}_{ij} = 0\bigr)} { M \sum_{i<j}
	\mathbb{I}\bigl(\{z_i^{\text{init}},z_j^{\text{init}}\}=\{k,l\}\bigr)\, 
	\mathbb{I}\bigl(A^{\text{init}}_{ij} = 0\bigr)},\\
	\beta_{kl}^{\text{init}} &= \frac{ \sum_{i<j}
	\mathbb{I}\bigl(\{z_i^{\text{init}},z_j^{\text{init}}\}=\{k,l\}\bigr)\, 
	\Bigl(M - \sum_{m=1}^M \tilde{A}^{(m)}_{ij}\Bigr)\, 
	\mathbb{I}\bigl(A^{\text{init}}_{ij} = 1\bigr)}{
	M \sum_{i<j} \mathbb{I}\bigl(\{z_i^{\text{init}},z_j^{\text{init}}\}=
	\{k,l\}\bigr)\, \mathbb{I}\bigl(A^{\text{init}}_{ij} = 1\bigr)}.
\end{align*}

\textit{(5) Initialization of covariate parameters}: We initialize the
parameters associated with $x^{\text{cont}}_{ir}$ using the sample means
and variances within each community defined by $\bm z^{\text{init}}$, which
are 
\begin{align*}
	\mu_{kr}^{\text{init}} = \frac{
	\sum_{i=1}^n \mathbb{I}\bigl(z_i^{\text{init}} = k\bigr)\, 
	x^{\text{cont}}_{ir}}{
	\sum_{i=1}^n \mathbb{I}\bigl(z_i^{\text{init}} = k\bigr)},\ 
	\tau_{kr}^{2,\text{init}}  = \frac{
	\sum_{i=1}^n \mathbb{I}\bigl(z_i^{\text{init}} = k\bigr)\,
	\bigl(x^{\text{cont}}_{ir} - \mu_{kr}^{\text{init}}\bigr)^2}{
	\sum_{i=1}^n \mathbb{I}\bigl(z_i^{\text{init}} = k\bigr)},\ k=1,\dots,K.
\end{align*}
Similarly, we initialize the parameters for categorical covariates as
\begin{align*}
	\xi_{ksc}^{\text{init}} &= \frac{ \sum_{i=1}^n 
	\mathbb{I}\bigl(z_i^{\text{init}} = k,\; x^\text{cat}_{i s} 
	= c\bigr) }{
	\sum_{i=1}^n  \mathbb{I}\bigl(z_i^{\text{init}} = k\bigr)},
	\quad c = 1,\dots,C_s.
\end{align*} 

\section*{Acknowledgements}
This work was supported in part by NIH Grant U01 AG068057.
This work was also supported by NIH Grants F31 AG091902 and
R01 AG071174. The content is solely the responsibility of the authors and does
not necessarily represent the official views of the National Institutes of Health.

The complete ADNI Acknowledgment is available at 
\url{https://adni.loni.usc.edu/wp-content/uploads/how_to_apply/
ADNI_Acknowledgement_List.pdf}. 

HCP data were provided by the Human Connectome Project, 
WU-Minn Consortium (Principal Investigators: David Van Essen and Kamil 
Ugurbil; 1U54MH091657) funded by the 16 NIH Institutes and Centers 
that support the NIH Blueprint for Neuroscience Research; and by 
the McDonnell Center for Systems Neuroscience at Washington University.
\vspace*{-8pt}

%

\section*{Data Availability}
The data that support the findings in this paper are available in the 
Alzheimer's Disease Neuroimaging Initiative database at 
\url{https://adni.loni.usc.edu} and the Human Connectome Project database 
at \url{http://www.humanconnectomeproject.org/data/}.

\bibliographystyle{chicago}
\bibliography{reference}

\end{document}